\documentclass{bmcart}

\usepackage{amsmath,amssymb}
\usepackage{changepage}
\usepackage[utf8x]{inputenc}
\usepackage{textcomp,marvosym}
\usepackage{cite}
\usepackage{nameref,hyperref}
\usepackage[right]{lineno}
\usepackage{microtype}
\DisableLigatures[f]{encoding = *, family = * }
\usepackage{array}

\newcolumntype{+}{!{\vrule width 2pt}}

\makeatletter
\renewcommand{\@biblabel}[1]{\quad#1.}
\makeatother

\usepackage{lastpage,fancyhdr,graphicx}
\usepackage{epstopdf}
\usepackage{multirow}
\usepackage{setspace}

\usepackage{color}
\usepackage{ulem}

\newcommand{\add}[1]{\textcolor{black}{#1}}
\newcommand{\del}[1]{\if0{#1}\fi}

\begin{document}

~\\
\doublespacing
Research \\
~\\
{\huge Detecting problematic transactions in a 
consumer-to-consumer e-commerce network} \\
~\\
Shun Kodate\textsuperscript{1,2}, 
Ryusuke Chiba\textsuperscript{3}, 
Shunya Kimura\textsuperscript{3}, 
Naoki Masuda\textsuperscript{4,5,2*} \\
~\\
* Corresponding author \\
naokimas@buffalo.edu \\

~\\
{\small
1 Graduate School of Information Sciences, Tohoku University, Sendai, 980-8579, Japan \\
2 Department of Engineering Mathematics, University of Bristol, Bristol, BS8 1UB, UK \\
3 Mercari, Inc., Tokyo, 106-6118, Japan \\
4 Department of Mathematics, University at Buffalo, Buffalo, NY, 14260-2900, USA \\
5 Computational and Data-Enabled Science and Engineering Program, University at Buffalo, Buffalo, NY, 14260-5030, USA \\
} \\

~\\
S.Kodate: kodate@sb.ecei.tohoku.ac.jp \\
RC: metalunk@mercari.com \\
S.Kimura: kimuras@mercari.com \\
NM: naokimas@buffalo.edu \\

\newpage

~\\
{\large Abstract} \\
Providers of online marketplaces are constantly combatting against 
problematic transactions, such as selling illegal items and posting 
fictive items, exercised by some of their users. 
A typical approach to detect fraud activity has been to analyze 
registered user profiles, user's behavior, and texts attached to 
individual transactions and the user. 
However, this traditional approach may be limited because  malicious 
users can easily conceal their information.
Given this background, network indices have been exploited for 
detecting frauds in various online transaction platforms.
In the present study, we analyzed networks of users of an online 
consumer-to-consumer marketplace in which a seller and the corresponding 
buyer of a transaction are connected by a directed edge. 
We constructed egocentric networks of each of several hundreds of 
fraudulent users and those of a similar number of normal users. 
We calculated eight local network indices based on up to connectivity
between the neighbors of the focal node. 
Based on the present descriptive analysis of these network indices, 
we fed twelve features that we constructed from the eight network indices
to random forest classifiers with the aim of distinguishing between 
normal users and fraudulent users engaged in each one of the four 
types of problematic transactions.
We found that the classifier accurately distinguished the fraudulent 
users from normal users and that the classification performance did not 
depend on the type of problematic transaction. 
\newpage

\doublespacing

\section*{Introduction}
In tandem with the rapid growth of online and electronic transactions 
and communications, fraud is expanding at a dramatic speed and 
penetrates our daily lives.
Fraud including cybercrimes costs billions of dollars per year and 
threatens the security of our society
\cite{commons2017, mcafee2018}.
In particular, in the recent era where online activity dominates, 
attacking a system is not too costly, whereas defending the system
against fraud is costly\cite{anderson2013}.
The dimension of fraud is vast and ranges from credit card fraud, money 
laundering, computer intrusion, to plagiarism, to name a few.

Computational and statistical methods for detecting and preventing 
fraud have been developed and implemented for decades
\cite{bolton2002, phua2010, abdallah2016, west2016}.
Standard practice for fraud detection is to employ statistical 
methods including the case of machine learning algorithms.
In particular, when both fraudulent and non-fraudulent samples are 
available, one can construct a classifier via supervised learning 
\cite{bolton2002, phua2010, abdallah2016, west2016}.
Exemplar features to be fed to such a statistical classifier include 
the transaction amount, day of the week, item category, and user's 
address for detecting frauds in credit card systems,
number of calls, call duration, call type, and user's age, gender, 
and geographical region in the case of telecommunication, and user 
profiles and transaction history in the case of online auctions
\cite{abdallah2016}.

However, many of these features can be easily faked by advanced 
fraudsters\cite{akoglu2015, google_whiteops2018}.
Furthermore, fraudulent users are adept at escaping the eyes of the 
administrators or authorities that would detect the usage of 
particular words as a signature of anomalous behavior
\cite{pu2006, hayes2007, bhowmick2016}.
For example, if the authority discovers that one jargon means a drug, 
then fraudulent users may easily switch to another jargon to confuse 
the authority.

Network analysis is an alternative way to construct features and is 
not new to fraud detection techniques\cite{savage2014, akoglu2015}. 
The idea is to use connectivity between nodes, which are usually
users or goods, in the given data and calculate graph-theoretic 
quantities or scores that characterize nodes.
These methods stand on the expectation that anomalous users show 
connectivity patterns that are distinct from those of normal users
\cite{akoglu2015}.
Network analysis has been deployed for fraud detection in 
insurance\cite{subelj2011}, 
money laundering\cite{drezewski2015, colladon2017, savage2017}, 
health-care data\cite{liu2016}, 
car-booking\cite{shchur2018}, 
a social security system\cite{vlasselaer2016}, 
mobile advertising\cite{hu2017}, 
a mobile phone network\cite{ferrara2014}, 
online social networks\cite{bhat2013, jiang2014, hooi2016, rasheed2018}, 
online review forums\cite{akoglu2013, liu2017, wang2018}, 
online auction or marketplaces\cite{chau2006, pandit2007, wang2008, 
bangcharoensap2015, yanchun2011}, 
credit card transactions\cite{vlasselaer2015, li2017}, 
cryptocurrency transaction\cite{monamo2016},
and various other fields\cite{akoglu2010}.
For example, fraudulent users and their accomplices were shown to form 
approximately bipartite cores in a network of users to inflate 
their reputations in an online auction system\cite{chau2006}.
Then, the authors proposed an algorithm based on a belief propagation 
to detect such suspicious connectivity patterns.
This method has been proven to be also effective on empirical data
obtained from eBay\cite{pandit2007}.

In the present study, we analyze a data set obtained from a large online 
consumer-to-consumer (C2C) marketplace, Mercari, operating in Japan and 
the US.
They are the largest C2C marketplace in Japan, in which, as of 2019, 
there are 13 million monthly active users and 133 billion yen 
(approximately 1.2 billion USD) transactions per quarter year
\cite{mercari_fact}.
Note that we analyze transaction frauds based on transaction networks of 
users, which contrasts with previous studies of online C2C marketplaces
that looked at reputation frauds
\cite{chau2006, pandit2007, wang2008, yanchun2011}.
Many prior network-based fraud detection algorithms used global 
information about networks, such as connected components, communities, 
betweenness, $k$-cores, and that determined by belief propagation
\cite{chau2006, pandit2007, wang2008, subelj2011, akoglu2013, bhat2013, 
ferrara2014, jiang2014, bangcharoensap2015, drezewski2015, vlasselaer2015, 
hooi2016, liu2016, vlasselaer2016, colladon2017, hu2017, li2017,
liu2017, savage2017, rasheed2018, shchur2018, wang2018}.
Others used local information about the users' network, such as 
the degree, the number of triangles, and the local clustering coefficient
\cite{chau2006, akoglu2010, subelj2011, yanchun2011, bangcharoensap2015, 
bhat2013, drezewski2015, monamo2016, vlasselaer2016, colladon2017}.
We will focus on local features of users, i.e., features of a node
that can be calculated from the connectivity of the user and the 
connectivity between neighbors of the user.
This is because local features are easier and faster to calculate and 
thus practical for commercial implementations.

\section*{Materials and methods}
\subsection*{Data}

Mercari is an online C2C marketplace service, where users trade various 
items among themselves.
The service is operating in Japan and the United States.
In the present study, we used the data obtained from the Japanese 
market \add{between July 2013 and January 2019}.
In addition to normal transactions, we focused on the following types 
of problematic transactions: 
fictive, underwear, medicine, and weapon. 
Fictive transactions are defined as selling non-existing items.
Underwear refers to transactions of used underwear; 
they are prohibited by the service from the perspective of morality and 
hygiene.
Medicine refers to transactions of medicinal supplies, which are 
prohibited by the law.
Weapon refers to transactions of weapons, which are prohibited by the 
service because they may lead to crime.
The number of sampled users of each type is shown in 
Table~\ref{table_summary}.

\subsection*{Network analysis}

\begin{figure}[bt]
\centering
\includegraphics[width=0.97\linewidth]{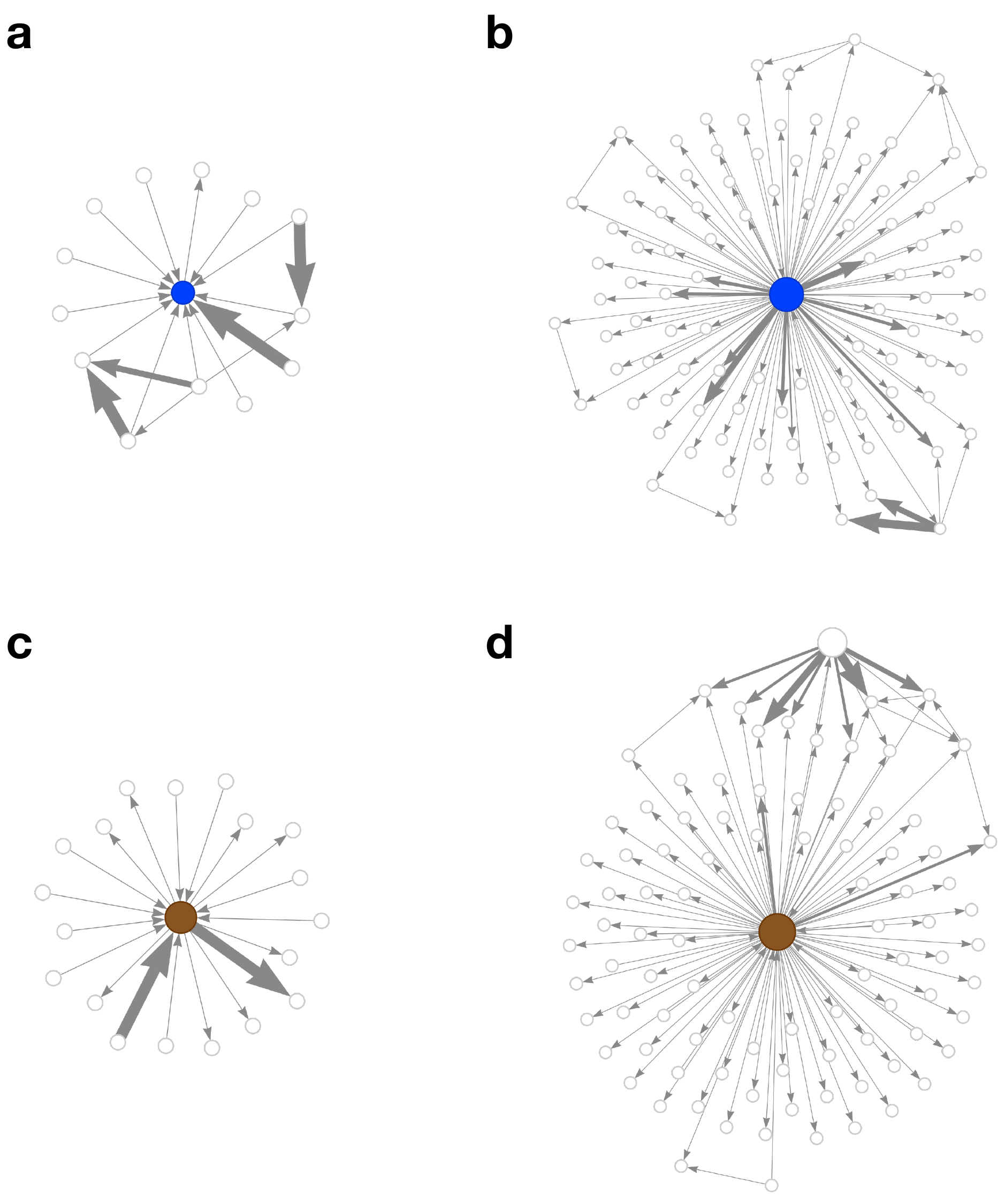}
\caption{\add{{\bf Examples of egocentric networks.}
(a) and (b): Egocentric networks of arbitrarily selected two normal users.
(c) and (d): Egocentric networks of arbitrarily selected two fraudulent users involved in 
selling a fictive item.
}}
\label{fig_egonet}
\end{figure}

We examine a directed and weighted network of users in which a user 
corresponds to a node and a transaction between two users
represents a directed edge.
The weight of the edge is equal to the number of transactions between 
the seller and the buyer.
We constructed egocentric networks of each of several hundreds of 
normal users and those of fraudulent users, i.e., those engaged in at 
least one problematic sell. 
\add{Figure~\ref{fig_egonet} shows the egocentric networks of two normal users 
(Fig.~\ref{fig_egonet}a, b)
and those of two fraudulent users involved in selling a fictive item 
(Fig.~\ref{fig_egonet}c, d).}
%
%
The egocentric network of either a normal or fraudulent user contained 
the nodes neighboring the focal user, edges between the focal user and 
these neighbors, and edges between the pairs of these neighbors. 

We calculated eight indices for each focal node. 
They are local indices in the meaning that they require the 
information up to the connectivity among the neighbors of the focal 
node.

Five out of the eight indices use only the information about the 
connectivity of the focal node.
The degree $k_i$ of node $v_i$ is the number of its neighbors.
The node strength~\cite{barrat2004architecture} (i.e., weighted degree)
of node $v_i$, denoted by $s_i$, is the number of transactions in which 
$v_i$ is involved.
Using these two indices, we also considered the mean number of 
transactions per neighbor, i.e., $s_i/k_i$, as a separate index.
These three indices do not use information about the direction of 
edges.

The sell probability of node $v_i$, denoted by $\mathrm{SP}_i$, 
uses the information about the direction of edges and defined as the 
proportion of the $v_i$'s neighbors for which $v_i$ acts as seller.
Precisely, the sell probability is given by 
\begin{eqnarray}
\label{eq:sellprob}
\mathrm{SP}_i = 
\frac{k_i^\mathrm{out}}{k_i^\mathrm{in}+k_i^\mathrm{out}},
\end{eqnarray}
where $k_i^\mathrm{in}$ is $v_i$'s in-degree (i.e., the number of 
neighbors from whom $v_i$ bought at least one item) and 
$k_i^\mathrm{out}$ is $v_i$'s out-degree (i.e., the number of 
neighbors to whom $v_i$ sold at least one item).
It should be noted that, if $v_i$ acted as both seller and buyer 
towards $v_j$, the contribution of $v_j$ to both in- and out-degree 
of $v_i$ is equal to one.
Therefore, $k_i^\mathrm{in} + k_i^\mathrm{out}$ is not equal to 
$k_i$ in general.

The weighted version of the sell probability, denoted by 
$\mathrm{WSP}_i$, is defined as
\begin{eqnarray}
\label{eq:w_sellprob}
\mathrm{WSP}_i = 
\frac{s_i^\mathrm{out}}{s_i^\mathrm{in}+s_i^\mathrm{out}},
\end{eqnarray}
where $s_i^\mathrm{in}$ is node $v_i$'s weighted in-degree (i.e., the 
number of buys) and $s_i^\mathrm{out}$ is $v_i$'s weighted 
out-degree (i.e., the number of sells).

The other three indices are based on triangles that involve the 
focal node.
The local clustering coefficient $C_i$ quantifies the abundance of 
undirected and unweighted triangles around $v_i$\cite{newman2010}.
It is defined as the number of undirected and unweighted triangles 
including $v_i$ divided by $k_i(k_i-1)/2$.
The local clustering coefficient $C_i$ ranges between 0 and 1.

We hypothesized that triangles contributing to an increase in the 
local clustering coefficient are localized around particular 
neighbors of node $v_i$.
Such neighbors together with $v_i$ may form an overlapping set 
of triangles, which may be regarded as a community 
\cite{radicchi2004defining, palla2005uncovering}.
Therefore, our hypothesis implies that the extent to which the focal 
node is involved in communities should be different between normal 
and fraudulent users.
To quantify this concept, we introduce the so-called triangle 
congregation, denoted by $m_i$. 
It is defined as the extent to which two triangles involving $v_i$ 
share another node and is given by
\begin{eqnarray}
\label{eq:m}
m_i = 
\frac{\textrm{(Number of pairs of triangles involving $v_i$ that 
share another node)}}
{\mathrm{Tr}_i(\mathrm{Tr}_i-1)/2},
\end{eqnarray}
where $\mathrm{Tr}_i = C_ik_i(k_i-1)/2$ is the number of 
triangles involving $v_i$.
Note that $m_i$ ranges between 0 and 1.

\begin{figure}[bt]
\centering
\includegraphics[width=0.97\linewidth]{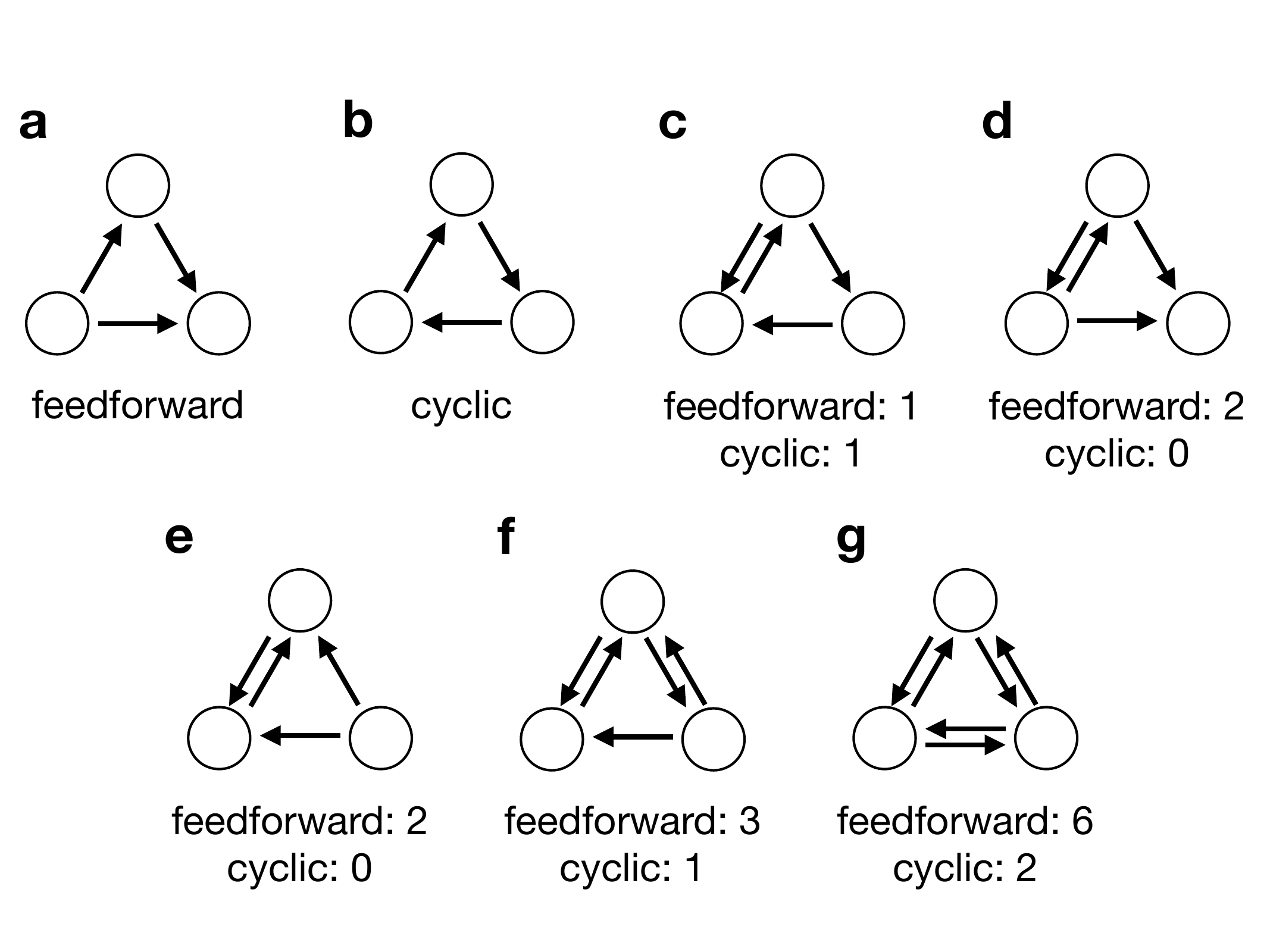}
\caption{{\bf Directed triangle patterns and their count.}
(a) Feedforward triangle. 
(b) Cyclic triangle. 
(c)--\del{(f) Four} \add{(g) Five} three-node patterns that contain directed triangles and 
reciprocal edges.
The numbers shown in the figure represent the number of feedforward 
or cyclic triangles to which each three-node pattern contributes.
}
\label{fig_motif}
\end{figure}

Frequencies of different directed three-node subnetworks, conventionally 
known as network motifs\cite{milo2002network}, may distinguish between 
normal and fraudulent users.
In particular, among triangles composed of directed edges, we 
hypothesized that feedforward triangles (Fig.~\ref{fig_motif}a) should 
be natural and that cyclic triangles (Fig.~\ref{fig_motif}b) are not.
We hypothesized so because a natural interpretation of a feedforward 
triangle is that a node with out-degree two tends to serve as seller 
while that with out-degree zero tends to serve as buyer and there are 
many such nodes that use the marketplace mostly as buyer or seller but 
not both. 
In contrast, an abundance of cyclic triangles may imply that 
relatively many users use the marketplace as both buyer and seller.
We used the index called the cycle probability, denoted by 
$\mathrm{CYP}_i$, which is defined by
\begin{eqnarray}
\label{eq:cyp}
\mathrm{CYP}_i = 
\frac {\mathrm{CY}_i} {\mathrm{FF}_i + \mathrm{CY}_i},
\end{eqnarray}
where $\mathrm{FF}_i$ and $\mathrm{CY}_i$ are the numbers of 
feedforward triangles and cyclic triangles to which node $v_i$ 
belongs. 
The definition of $\mathrm{FF}_i$ and $\mathrm{CY}_i$, and hence 
$\mathrm{CYP}_i$, is valid even when the triangles involving $v_i$
have bidirectional edges.
In the case of Fig.~\ref{fig_motif}c, for example, any of the three 
nodes contains one feedforward triangle and one cyclic triangle.
The other four cases in which bidirectional edges are involved in 
triangles are shown in Figs.~\ref{fig_motif}d-\del{f}\add{g}.
In the calculation of $\mathrm{CYP}_i$, we ignored the weights of edges.


\subsection*{Random forest classifier}
To classify users into normal and fraudulent users based on their local 
network properties, we employed a random forest classifier
\cite{breiman2001random, breiman1984classification, hastie2009elements} 
implemented in scikit-learn\cite{pedregosa2011scikit}.
It uses an ensemble learning method that combines multiple classifiers, 
each of which is a decision tree, built from training data and 
classifies test data avoiding overfitting.
We combined 300 decision-tree classifiers to construct a random 
forest classifier.
Each decision tree is constructed on the basis of training samples 
that are randomly subsampled with replacement from the set of all the 
training samples.
To compute the best split of each node in a tree, one randomly samples 
the candidate features from the set of all the features.
The probability that a test sample is positive in a tree is estimated as 
follows.
Consider the terminal node in the tree that a test sample eventually 
reaches.
The fraction of positive training samples at the terminal node gives
the probability that the test sample is classified as positive.
One minus the positive probability gives the negative probability 
estimated for the same test sample.
The positive or negative probability for the random forest classifier 
is obtained as the average of single-tree positive or negative 
probability over all the 300 trees.
A sample is classified as positive by the random forest classifier if the
positive probability is larger than 0.5, otherwise classified as negative.

\add{We split samples of each type into two sets such that 75\% and 25\% 
of the samples of each type are assigned to the training and test samples, 
respectively.}
There were more normal users than any \del{one} type of fraudulent user.
Therefore, to balance the number of the negative (i.e., normal) and 
positive (i.e., fraudulent) samples, we uniformly randomly subsampled the 
negative samples \add{(i.e., under-sampling)} 
such that the number of the samples is the same between 
the normal and fraudulent types \add{in the training set}.
\del{Then, we split samples of each type into two sets such that 75\% and 25\% 
of the} \\
\del{samples of each type are assigned to the training and test samples, 
respectively.} \\
Based on the training sample constructed in this manner, we built each 
of the 300 decision trees and hence a random forest classifier.
Then, we examined the classification performance of the random forest 
classifier on the set of test samples.

The true positive rate, also called the recall, is defined as the 
proportion of the positive samples (i.e., fraudulent users) that 
the random forest classifier correctly classifies as positive.
The false positive rate is defined as the proportion of the negative
samples (i.e., normal users) that are incorrectly classified as positive.
The precision is defined as the proportion of the truly positive samples 
among those that are classified as positive.
The true positive rate, false positive rate, and precision range between 
0 and 1.

We used the following two performance measures for the random forest 
classifier.
To draw the receiver operating characteristic (ROC) curve for a random 
forest classifier, one first arranges the test samples in descending 
order of the estimated probability that they are positive.
Then, one plots each test sample, with its false positive rate on the 
horizontal axis and the true positive rate on the vertical axis.
By connecting the test samples in a piecewise linear manner, one 
obtains the ROC curve.
The precision-recall (PR) curve is generated by plotting the samples 
in the same order in $[0, 1]^2$, with the recall on the horizontal axis
and the precision on the vertical axis.
For an accurate binary classifier, both ROC and PR curves visit near 
$(x, y) = (0, 1)$. 
Therefore, we quantify the performance of the classifier by the area
under the curve (AUC) of each curve.
The AUC ranges between 0 and 1, and a large value indicates a good 
performance of the random forest classifier.

To calculate the importance of each feature in the random forest 
classifier, we used the permutation importance
\cite{strobl2007bias, altmann2010permutation}.
With this method, the importance of a feature is given by the decrease
in the performance of the trained classifier when the feature is randomly
permuted among the test samples.
A large value indicates that the feature considerably contributes to 
the performance of the classifier.
To calculate the permutation importance, we used the AUC value of the 
ROC curve as the performance measure of a random forest classifier.
We computed the permutation importance of each feature with ten 
different permutations and adopted the average over the ten permutations 
as the importance of the feature.

We optimized the parameters of the random forest classifier by a grid 
search with 10-fold cross-validation on the training set.
For the maximum depth of each tree (i.e., the max\_depth parameter in 
scikit-learn), we explored the integers between 3 and 10.
For the number of candidate features for each split (i.e., max\_features), 
we explored the integers between 3 and 6.
For the minimum number of samples required at terminal nodes 
(i.e., min\_samples\_leaf), we explored 1, 3, and 5.
As mentioned above, the number of trees (i.e., n\_estimators) was set to 
300.
The seed number for the random number generator (i.e., random\_state) was 
set to 0.
For the other hyperparameters, we used the default values in scikit-learn 
version 0.22.
In the parameter optimization, we evaluated the performance of the 
random forest classifier with the AUC value of the ROC curve measured
on a single set of training and test samples.

To avoid sampling bias, we built 100 random forest classifiers, trained
each classifier, and tested its performance on a randomly drawn set of 
train and test samples, whose sampling scheme was described above.

\section*{Results}
\subsection*{Descriptive statistics}

\begin{figure}[bt]
\centering
\includegraphics[width=0.97\linewidth]{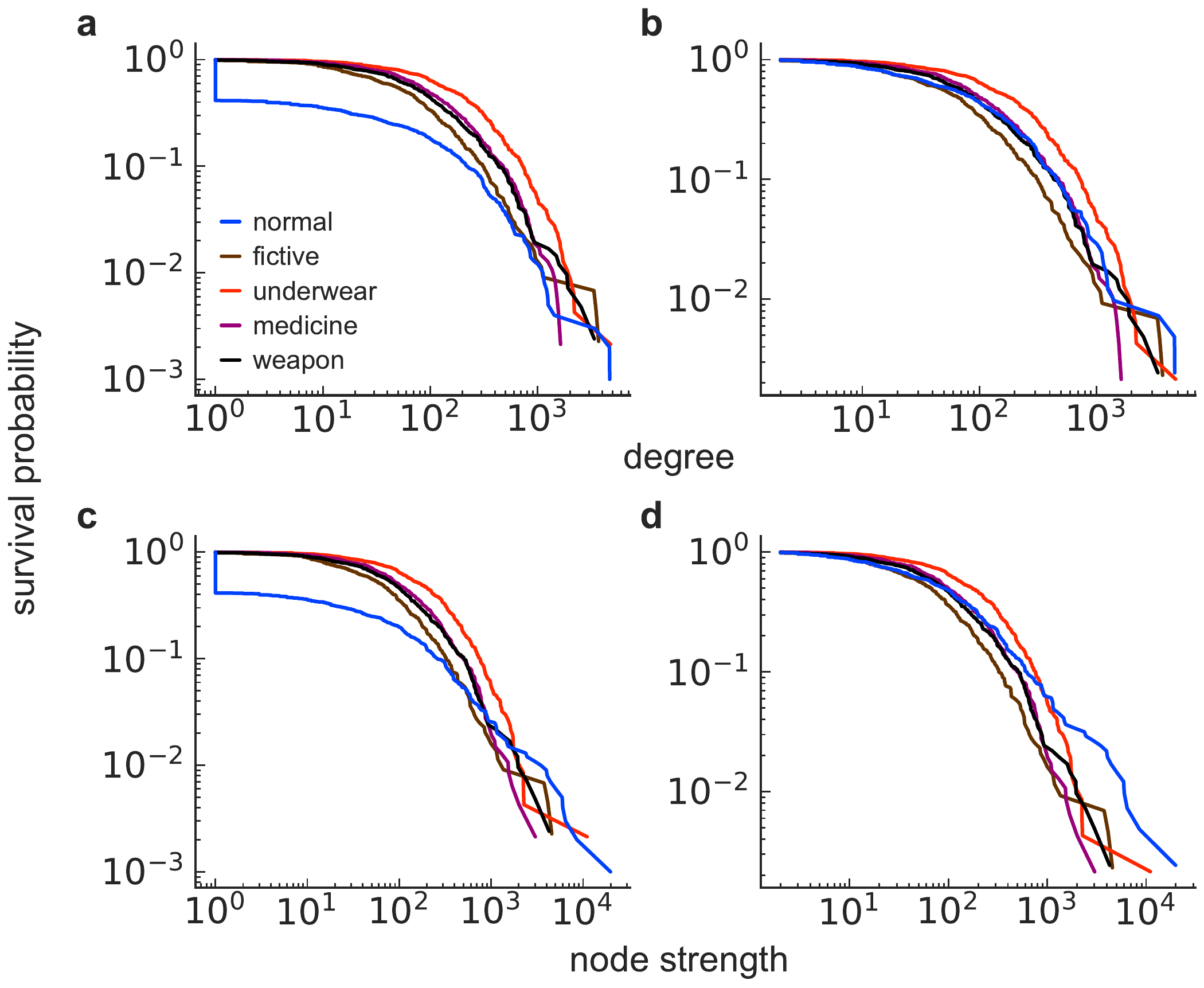}
\caption{{\bf Survival probability of the degree for each user type.}
(a) Degree (i.e., $k_i$) for all nodes.
(b) Degree for the nodes with $k_i\ge2$.
(c) Strength (i.e., $s_i$) for all nodes.
(d) Strength for the nodes with $s_i\ge2$.
}
\label{fig_degree}
\end{figure}

The survival probability of the degree (i.e., a fraction of nodes whose 
degree is larger than a specified value) is shown in 
Fig.~\ref{fig_degree}a for each user type. 
Approximately 60\% of the normal users have degree $k_i=1$, whereas the
fraction of the users with $k_i=1$ is approximately equal to 2\% or less 
for any type of fraudulent user (Table~\ref{table_summary}).
Therefore, we expect that whether $k_i=1$ or $k_i\ge2$ gives useful
information for distinguishing between normal and fraudulent users.
The degree distribution at $k_i\ge2$ may provide further information 
useful for the classification.
The survival probability of the degree distribution conditioned on 
$k_i\ge2$ for the different types of users is shown in 
Fig.~\ref{fig_degree}b.
The figure suggests that the degree distribution is systematically 
different between the normal and fraudulent users.
However, we consider that the difference is not as clear-cut as that 
in the fraction of users having $k_i=1$ (Table~\ref{table_summary}).

\newcolumntype{L}[1]{>{\raggedright\let\newline\\\arraybackslash\hspace{0pt}}m{#1}}
\newcolumntype{C}[1]{>{\centering\let\newline\\\arraybackslash\hspace{0pt}}m{#1}}
\newcolumntype{R}[1]{>{\raggedleft\let\newline\\\arraybackslash\hspace{0pt}}m{#1}}

\begin{table}[!b]
\begin{adjustwidth}{-0.85in}{0in}
\centering
\caption{{\bf Properties of different types of users.}
In the first column, mean($\mathrm{A}\mid\mathrm{B}$), for example, 
represents the mean of A conditioned on B.
Unless the first column mentions the conditional mean, median, 
\add{or the number of transactions}, the numbers reported in the table represent 
the number of users.
}
\begin{tabular}{C{4.1cm}C{2.2cm}C{2.2cm}C{2.2cm}C{2.2cm}C{2.2cm}}
Seed user type & Normal & Fictive & Underwear & Medicine & Weapon \\
\hline
Number of seed users & 999 & 440 & 468 & 469 & 416 \\
\hline
\add{Number of transactions involving the seed user} & \add{151,021} & \add{66,215} & \add{151,278} & \add{92,497} & \add{81,970} \\
\add{Total number of transactions} & \add{27,683,860} & \add{850,739} & \add{2,325,898} & \add{925,361} & \add{533,963} \\
\hline
$k_i=1$ & 587 (58.8\%) & 8 (1.8\%) & 3 (0.6\%) & 2 (0.4\%) & 5 (1.2\%) \\
mean($k_i \mid k_i\ge2$) & 195.0 & 138.3 & 297.8 & 184.2 & 179.7 \\
median($k_i \mid k_i\ge2$) & 77.5 & 61.0 & 170.0 & 97.0 & 86.0 \\
\hline
$s_i=1$ & 587 (58.8\%) & 8 (1.8\%) & 3 (0.6\%) & 2 (0.4\%) & 5 (1.2\%) \\
mean($s_i \mid s_i\ge2$) & 365.1 & 153.3 & 325.3 & 198.1 & 199.4 \\
median($s_i \mid s_i\ge2$) & 89.0 & 66.5 & 175.0 & 100.0 & 90.0 \\
\hline
$s_i\ge2$ & 412 & 432 & 465 & 467 & 411 \\
$s_i/k_i=1$ & 97 (23.5\%) & 97 (22.5\%) & 86 (18.5\%) & 156 (33.4\%) & 121 (29.4\%) \\
mean($s_i/k_i \mid s_i/k_i>1$) & 1.413 & 1.135 & 1.055 & 1.066 & 1.092 \\
median($s_i/k_i \mid s_i/k_i>1$) & 1.124 & 1.059 & 1.03 & 1.031 & 1.055 \\
\hline
$k_i\ge2$ & 412 & 432 & 465 & 467 & 411 \\
$\mathrm{SP}_i=1$ & 157 (38.1\%) & 15 (3.5\%) & 21 (4.5\%) & 16 (3.4\%) & 17 (4.1\%) \\
$k_i^\mathrm{out}=1$ & 118 (28.6\%) & 21 (4.9\%) & 2 (0.4\%) & 2 (0.4\%) & 9 (2.2\%) \\
\hline
$s_i\ge2$ & 412 & 432 & 465 & 467 & 411 \\
$\mathrm{WSP}_i=1$ & 157 (38.1\%) & 15 (3.5\%) & 21 (4.5\%) & 16 (3.4\%) & 17 (4.1\%) \\
$s_i^\mathrm{out}=1$ & 118 (28.6\%) & 14 (3.2\%) & 2 (0.4\%) & 2 (0.4\%) & 9 (2.2\%) \\
\hline
$k_i\ge2$ & 412 & 432 & 465 & 467 & 411 \\
$C_i=0$ & 118 (28.6\%) & 152 (35.2\%) & 108 (23.2\%) & 154 (33.0\%) & 128 (31.1\%) \\
mean($C_i \mid C_i>0$) & $8.554\times10^{-3}$ & $8.348\times10^{-3}$ & $9.500\times10^{-4}$ & $2.231\times10^{-3}$ & $3.810\times10^{-3}$ \\
median($C_i \mid C_i>0$) & $2.411\times10^{-3}$ & $2.039\times10^{-3}$ & $5.288\times10^{-4}$ & $6.494\times10^{-4}$ & $1.337\times10^{-3}$ \\
\hline
$\mathrm{Tr}_i\ge2$ & 262 & 241 & 317 & 251 & 244 \\
$m_i=0$ & 17 (6.5\%) & 27 (11.2\%) & 54 (17.0\%) & 44 (17.5\%) & 32 (13.1\%) \\
$m_i=1$ & 12 (4.6\%) & 9 (3.7\%) & 4 (1.3\%) & 6 (2.4\%) & 11 (4.5\%) \\
mean($m_i \mid m_i>0$) & $8.554\times10^{-3}$ & $8.348\times10^{-3}$ & $9.500\times10^{-4}$ & $2.231\times10^{-3}$ & $3.810\times10^{-3}$ \\
median($m_i \mid m_i>0$) & $2.411\times10^{-3}$ & $2.039\times10^{-3}$ & $5.288\times10^{-4}$ & $6.494\times10^{-4}$ & $1.337\times10^{-3}$ \\
\hline
$\mathrm{FF}_i+\mathrm{CY}_i\ge1$ & 294 & 280 & 357 & 313 & 283 \\
$\mathrm{CYP}_i=0$ & 234 (79.6\%) & 188 (67.1\%) & 222 (62.2\%) & 227 (72.5\%) & 202 (71.4\%) \\
mean($\mathrm{CYP}_i \mid \mathrm{CYP}_i>0$) & $1.987\times10^{-2}$ & $7.367\times10^{-2}$ & $6.739\times10^{-2}$ & $8.551\times10^{-2}$ & $5.544\times10^{-2}$ \\
median($\mathrm{CYP}_i \mid \mathrm{CYP}_i>0$) & $1.521\times10^{-2}$ & $4.481\times10^{-2}$ & $3.396\times10^{-2}$ & $3.822\times10^{-2}$ & $3.618\times10^{-2}$ \\
\hline
\end{tabular}
\label{table_summary}
\end{adjustwidth}
\end{table}


The survival probability of the node strength (i.e., weighted degree) 
is shown in Fig.~\ref{fig_degree}c for each user type.
As in the case for the unweighted degree, we found that many normal users, 
but not fraudulent users, have $s_i=1$.
In fact, the number of the normal users with $s_i=1$ is equal to those with
$k_i=1$ (Table~\ref{table_summary}), implying that all normal users with 
$k_i=1$ participated in just one transaction.
In contrast, no user had $s_i=1$ for any type of fraudulent user.
The survival probability of the node strength conditioned on 
$s_i\ge2$ apparently does not show a clear distinction between the 
normal and fraudulent users 
(Fig.~\ref{fig_degree}d, Table~\ref{table_summary}).

\begin{figure}[bt]
\centering
\includegraphics[width=0.97\linewidth]{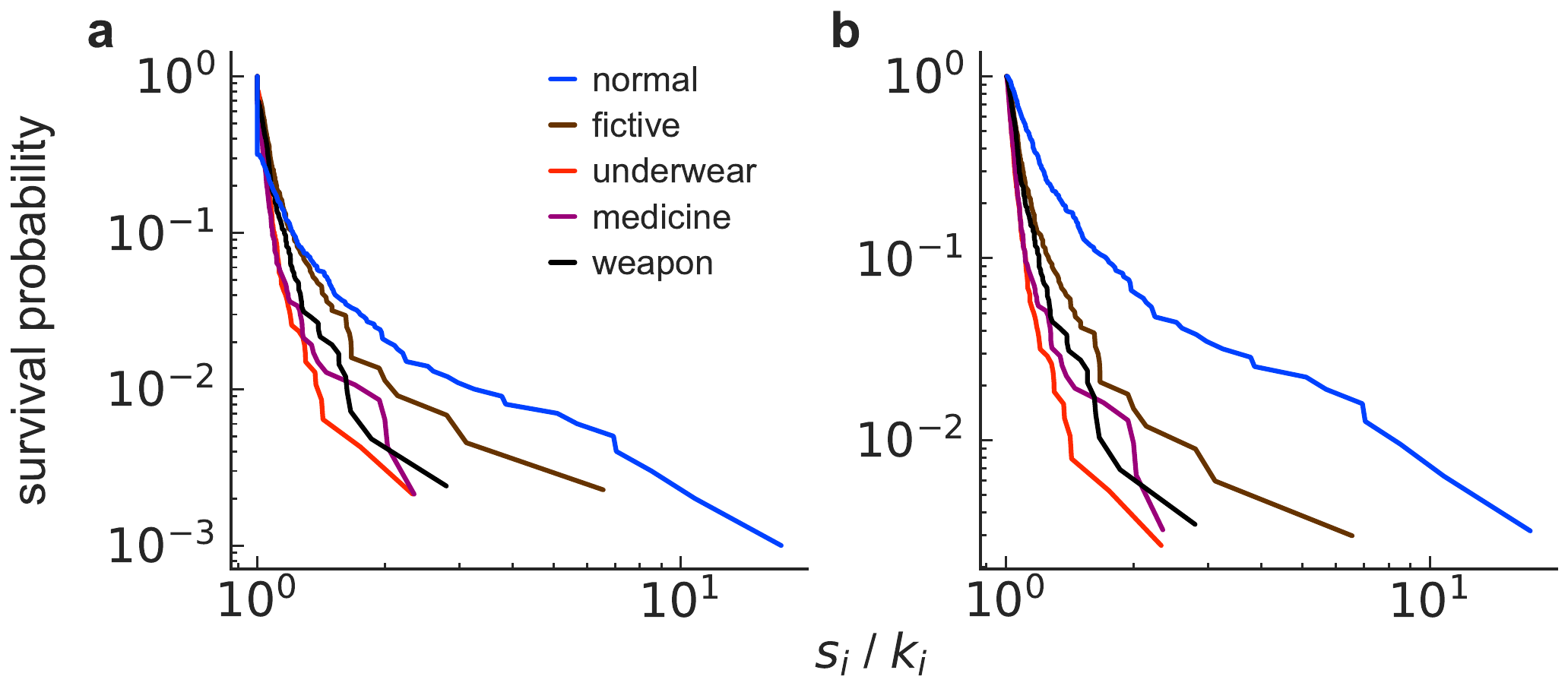}
\caption{{\bf Survival probability of the average number of transactions
per neighbor.}
(a) $s_i/k_i$ for all nodes.
(b) $s_i/k_i$ for the nodes with $s_i/k_i>1$.
}
\label{fig_spk}
\end{figure}

The distribution of the average number of transactions per edge, i.e.,
$s_i/k_i$, is shown in Fig.~\ref{fig_spk}a.
We found that a majority of normal users have $s_i/k_i=1$.
This result indicates that a large fraction of normal users is engaged in 
just one transaction per neighbor (Table~\ref{table_summary}).
This result is consistent with the fact that approximately 60\% of the 
normal users have $k_i=s_i=1$.
In contrast, many of any type of fraudulent users have $s_i/k_i>1$.
However, they tend to have a smaller value of $s_i/k_i$ than the normal users.
This difference is more noticeable when we discraded the users with $s_i/k_i=1$
(Fig.~\ref{fig_spk}b, Table~\ref{table_summary}).
Therefore, less frequent transactions with a specific neighbor seem to be a 
characteristic behavior of fraudulent users.

\begin{figure}[bt]
\centering
\includegraphics[width=0.97\linewidth]{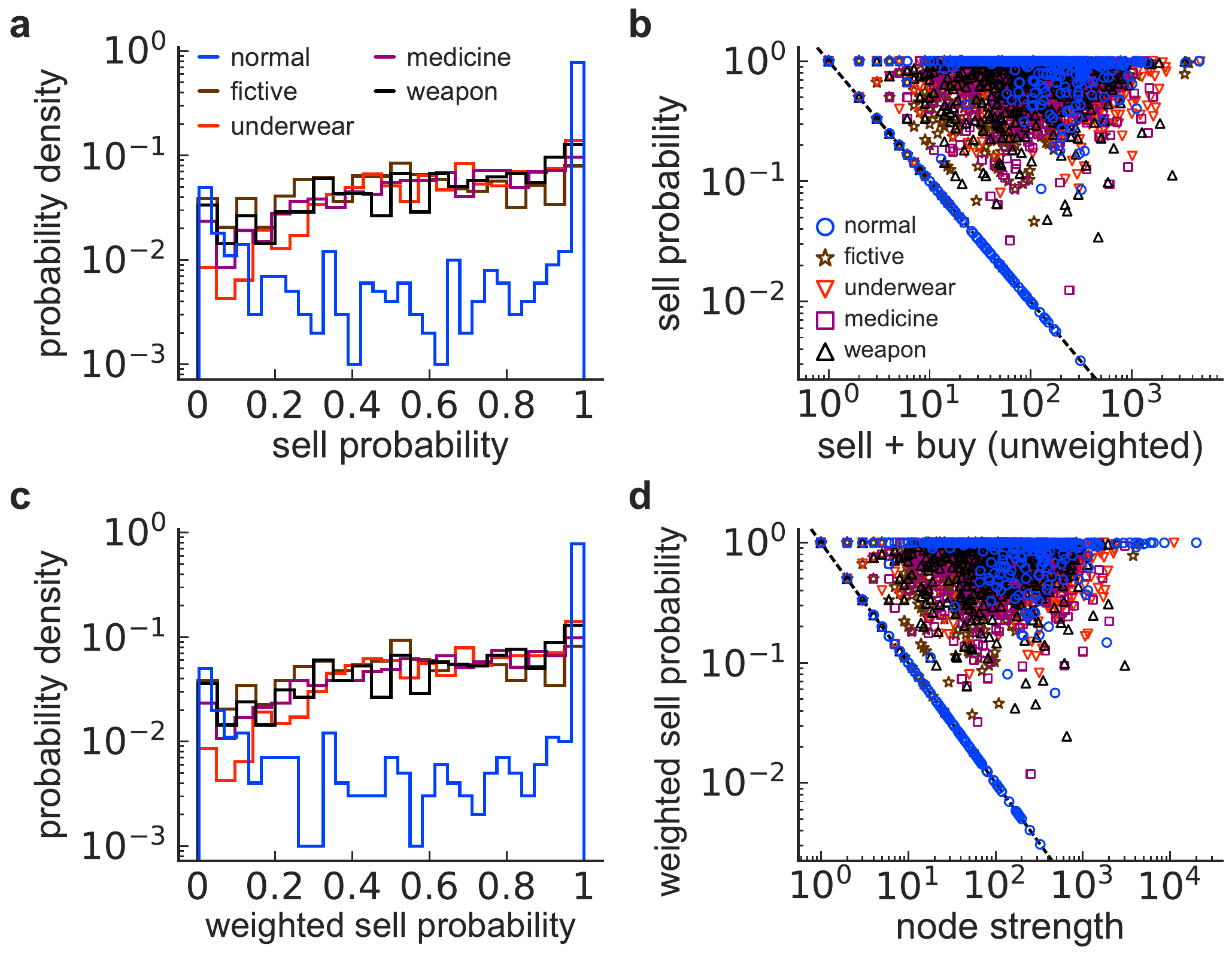}
\caption{{\bf Sell probability for each user type.}
(a) Distribution of the unweighted sell probability.
(b) Relationship between the degree and the unweighted sell 
probability.
(c) Distribution of the weighted sell probability.
(d) Relationship between the node strength and the weighted sell 
probability.
The dashed lines in (b) and (d) indicate $1/(k_i^\mathrm{in}+k_i^\mathrm{out})$ 
and $1/(s_i^\mathrm{in}+s_i^\mathrm{out})$, respectively.
}
\label{fig_sp}
\end{figure}

The distribution of the unweighted sell probability for the 
different user types is shown in Fig.~\ref{fig_sp}a.
The distribution for the normal users is peaked around 0 and 1, 
indicating that a relatively large fraction of normal users is almost
exclusive buyer or seller. 
Note that, by definition, the sell probability is at least 
$1/(k_i^\mathrm{in}+k_i^\mathrm{out})$ because our samples are sellers.
Therefore, a peak around the sell probability of zero implies that 
the users probably have no or few sell transactions apart from the one 
sell transaction based on which the users have been sampled as seller.
In contrast, the distribution for any fraudulent type is relatively flat.
Figure~\ref{fig_sp}b shows the relationships between the unweighted 
sell probability and the degree.
On the dashed line in Fig.~\ref{fig_sp}b, the sell probability is 
equal to $1/(k_i^\mathrm{in}+k_i^\mathrm{out})$, indicating that the node 
has $k_i^\mathrm{out}=1$, which is the smallest possible out-degree.
The users on this line were buyers in all but one transaction.
Figure~\ref{fig_sp}b indicates that a majority of such users are 
normal as opposed to fraudulent users, which is quantitatively confirmed
in Table~\ref{table_summary}.
We also found that most of the normal users were either on the 
horizontal line with the sell probability of one (38.1\% of the 
normal users with $k_i\ge2$; see Table~\ref{table_summary} for the 
corresponding fractions of normal users with $k_i=1$) or on the dashed 
line (28.6\%).
This is not the case for any type of fraudulent user 
(Table~\ref{table_summary}).

The distribution of the weighted sell probability for the different 
user types and the relationships between the weighted sell probability 
and the node strength are shown in Fig.~\ref{fig_sp}c and 
Fig.~\ref{fig_sp}d, respectively.
The results are similar to the case of the unweighted sell probability in 
two aspects.
First, the normal users and the fraudulent users form distinct frequency 
distributions (Fig.~\ref{fig_sp}c).
Second, most of the normal users are either on the horizontal line with 
the weighted sell probability of one or on the dashed line with the 
smallest possible weighted sell probability, i.e., $1/s_i$ 
(Fig.~\ref{fig_sp}d, Table~\ref{table_summary}).

\begin{figure}[bt]
\centering
\includegraphics[width=0.97\linewidth]{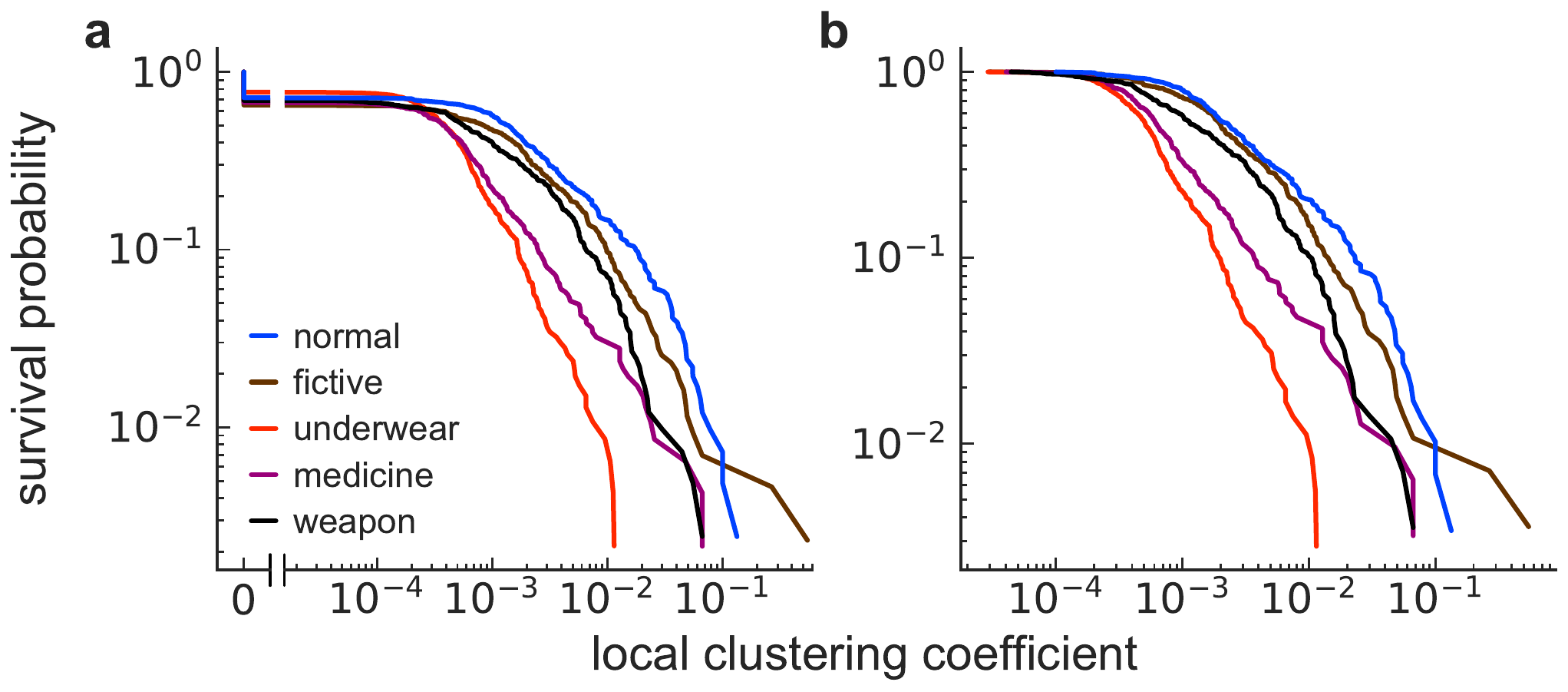}
\caption{
{\bf Local clustering coefficient for each user type.}
(a) Survival probability.
(b) Survival probability conditioned on $C_i>0$.
}
\label{fig_clst}
\end{figure}

The survival probability of the local clustering coefficient is shown 
in Fig.~\ref{fig_clst}a. 
It should be noted that, in this analysis, we confined ourselves to 
the users with $k_i\ge2$ because $C_i$ is undefined when $k_i=1$.
We found that the number of users with $C_i=0$ is not considerably 
different between the normal and fraudulent users 
(also see Table~\ref{table_summary}).
Figure~\ref{fig_clst}b shows the survival probability of $C_i$
conditioned on $C_i>0$.
The normal users tend to have a larger value of $C_i$ than fraudulent 
users, whereas this tendency is not strong (Table~\ref{table_summary}).

\begin{figure}[bt]
\centering
\includegraphics[width=0.97\linewidth]{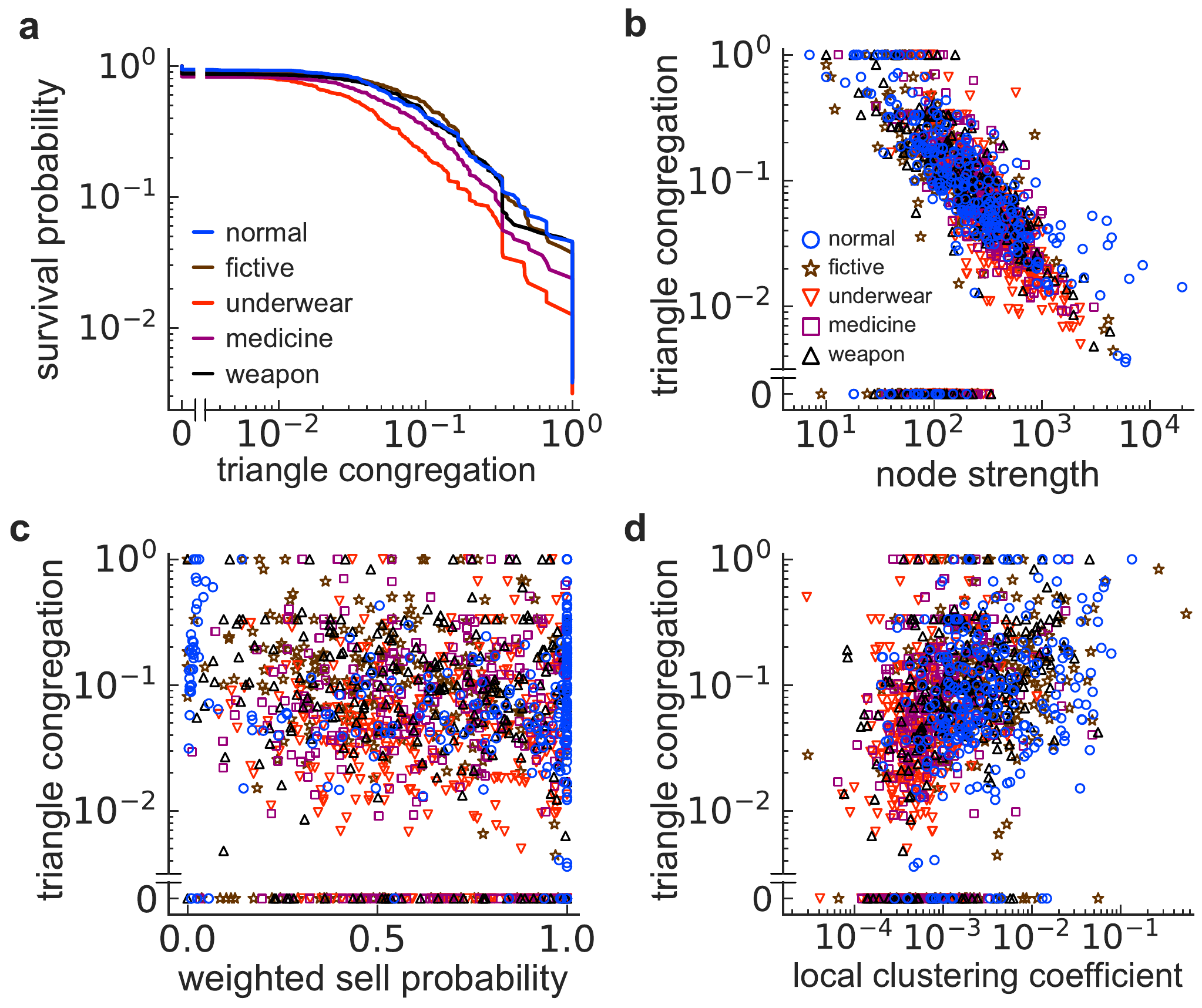}
\caption{{\bf Triangle congregation for each user type.}
(a) Survival probability.
(b) Relationship between the triangle congregation, $m_i$, and 
the node strength.
(c) Relationship between $m_i$ and the weighted sell probability.
(d) Relationship between $m_i$ and the local clustering coefficient.
}
\label{fig_m}
\end{figure}

The survival probability of the triangle congregation is shown in 
Fig.~\ref{fig_m}a.
Contrary to our hypothesis, there is no clear difference between the
distribution of the normal and fraudulent users.
The triangle congregation tends to be large when the node strength 
is small (Fig.~\ref{fig_m}b) and the local clustering coefficient is 
large (Fig.~\ref{fig_m}d).
It depends little on the weighted sell probability (Fig.~\ref{fig_m}c).
However, we did not find clear differences in the triangle 
congregation between the normal and fraudulent users (also see 
Table~\ref{table_summary}).

\begin{figure}[bt]
\centering
\includegraphics[width=0.97\linewidth]{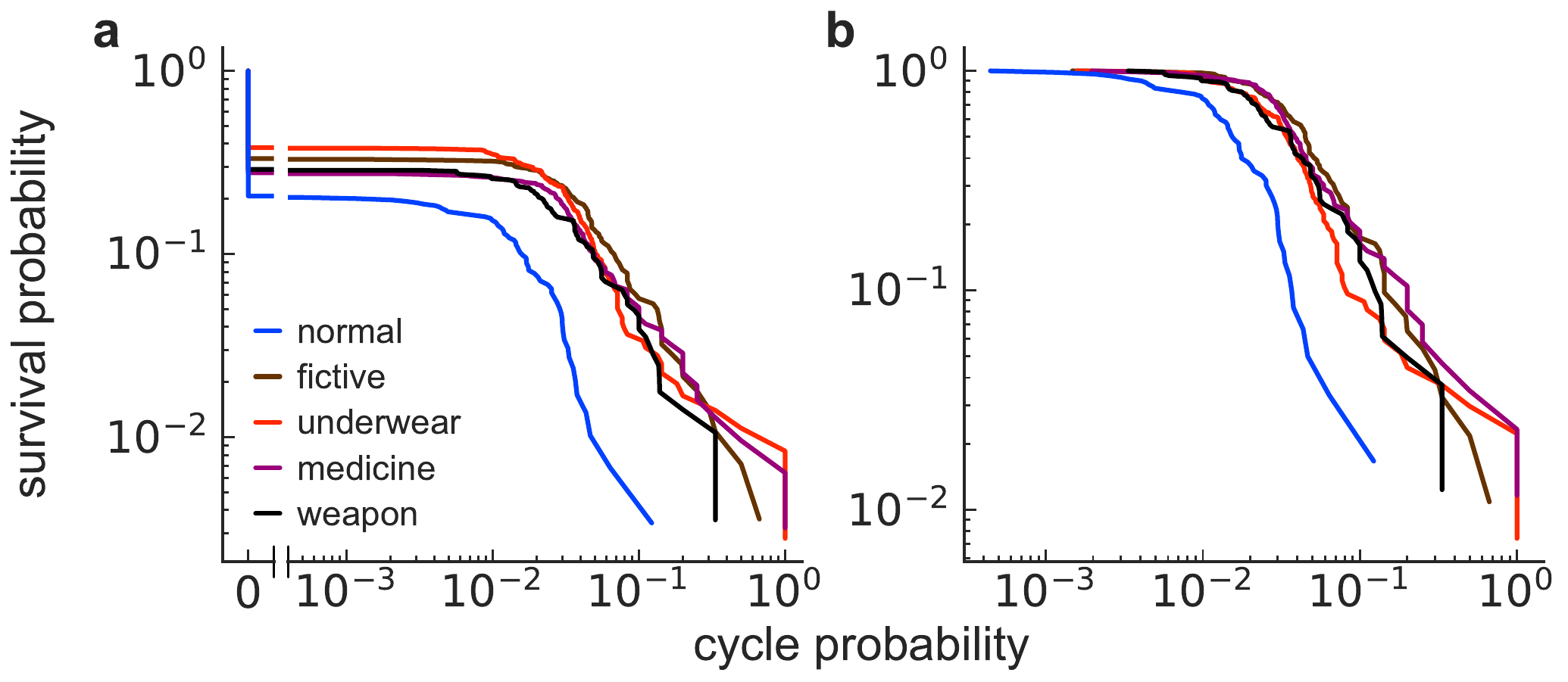}
\caption{{\bf Cycle probability for each user type.}
(a) Survival probability.
(b) Survival probability conditioned on CYP$_i>0$.
}
\label{fig_cyp}
\end{figure}

The survival probability of the cycle probability is shown in 
Fig.~\ref{fig_cyp}a.
A large fraction of any type of users has $\mathrm{CYP}_i=0$ 
(Table~\ref{table_summary}).
When the users with $\mathrm{CYP}_i=0$ are discarded, the normal users
tend to have a smaller value of $\mathrm{CYP}_i$ than any type of fraudulent 
users (Fig.~\ref{fig_cyp}b, Table~\ref{table_summary}).


\subsection*{Classification of users}
Based on the eight indices whose descriptive statistics were analyzed 
in the previous section, we defined 12 features and fed them to the 
random forest classifier.
The aim of the classifier is to distinguish between normal and fraudulent 
users.
The first feature is binary and whether the degree $k_i=1$ or 
$k_i\ge2$.
The second feature is also binary and whether the node strength 
$s_i=1$ or $s_i\ge2$.
The third feature is $s_i/k_i$, which is a real number greater than 
or equal to 1.
The fourth feature is binary and whether the unweighted sell 
probability $\mathrm{SP}_i=1$ or $\mathrm{SP}_i<1$.
The fifth feature is binary and whether 
$\mathrm{SP}_i=1/(k_i^\mathrm{in}+k_i^\mathrm{out})$ or  
$\mathrm{SP}_i>1/(k_i^\mathrm{in}+k_i^\mathrm{out})$, i.e., 
whether $k_i^\mathrm{out}=1$ or $k_i^\mathrm{out}>1$.
The sixth feature is $\mathrm{SP}_i$, which ranges between 0 and 1.
The seventh feature is binary and whether the weighted sell 
probability $\mathrm{WSP}_i=1$ or $\mathrm{WSP}_i<1$.
The eighth feature is binary and whether 
$\mathrm{WSP}_i=1/(s_i^\mathrm{in}+s_i^\mathrm{out})$ or  
$\mathrm{WSP}_i>1/(s_i^\mathrm{in}+s_i^\mathrm{out})$, i.e., 
whether $s_i^\mathrm{out}=1$ or $s_i^\mathrm{out}>1$.
The ninth feature is $\mathrm{WSP}_i$, which ranges between 0 and 1. 
The tenth feature is the local clustering coefficient $C_i$, which 
ranges between 0 and 1.
When $k_i=1$, the local clustering coefficient is undefined.
In this case, we set $C_i=-1$.
The eleventh feature is the triangle congregation $m_i$, which ranges
between 0 and 1.
When there is no triangle or only one triangle involving $v_i$, one cannot 
calculate $m_i$.
In this case, we set $m_i=-1$.
Finally, the twelfth feature is the cycle probability 
$\mathrm{CYP}_i$, which ranges between 0 and 1.
When there is neither feedforward nor cyclic triangle involving $v_i$, 
$\mathrm{CYP}_i$ is undefined.
In this case, we set $\mathrm{CYP}_i=-1$.

\begin{figure}[bt]
\centering
\includegraphics[width=0.97\linewidth]{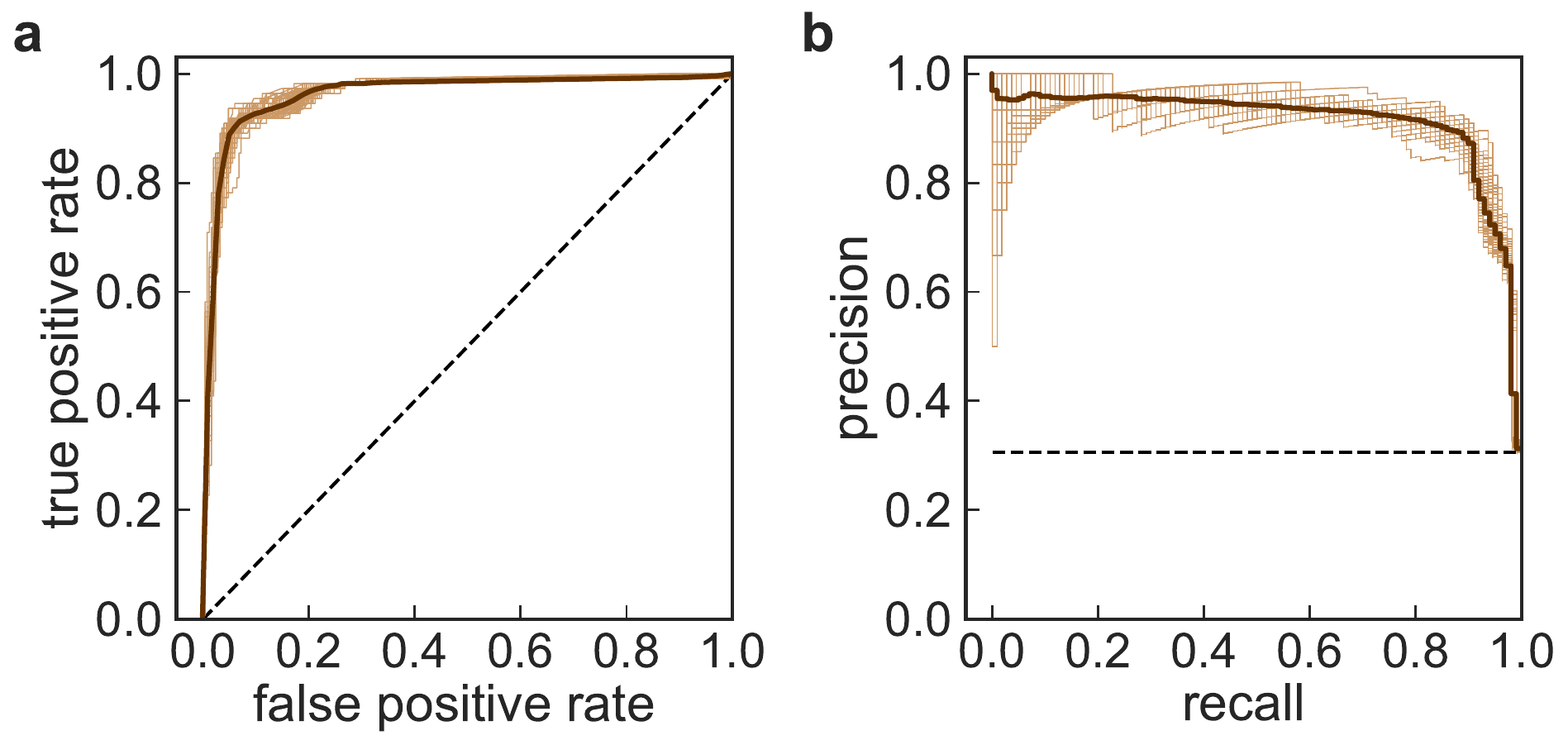}
\caption{{\bf ROC and PR curves when the normal users and those involved 
in fictive transactions are classified.}
(a) ROC curves.
(b) PR curves.
Each thin line corresponds to one of the 100 classifiers.
The thick lines correspond to the average of the 100 lines.
The dashed lines correspond to the uniformly random classification.
}
\label{fig_roc_pr}
\end{figure}

The ROC and PR curves when all the 12 features of users are used and the 
fraudulent type is fictive transactions are shown in 
Figs.~\ref{fig_roc_pr}a and b, respectively.
Each thin line corresponds to one of the 100 classifiers.
The thick lines correspond to the average of the 100 lines.
The dashed lines correspond to the uniformly random classification.
Figure~\ref{fig_roc_pr} indicates that the classification performance 
seems to be high.
Quantitatively, for this and the other types of fraudulent users, the AUC
values always exceeded \del{0.94} \add{0.91} (Table~\ref{table_auc}).


\begin{table}[bt]
\centering
\caption{{\bf AUC values for the random forest classifiers.}
The average and standard deviation were calculated based on the 100 
classifiers.}
\begin{tabular}{rrC{2.08cm}C{2.08cm}C{2.08cm}C{2.08cm}}
& & Fictive & Underwear & Medicine & Weapon \\
\hline
\multirow{2}{*}{12 features}
& ROC & \del{0.955 $\pm$ 0.011} \add{0.962 $\pm$ 0.003} & \del{0.978 $\pm$ 0.005} \add{0.981 $\pm$ 0.001} & \del{0.967 $\pm$ 0.009} \add{0.979 $\pm$ 0.003} & \del{0.963 $\pm$ 0.010} \add{0.969 $\pm$ 0.004} \\
& PR & \del{0.949 $\pm$ 0.015} \add{0.916 $\pm$ 0.009} & \del{0.978 $\pm$ 0.010} \add{0.948 $\pm$ 0.006} & \del{0.967 $\pm$ 0.011} \add{0.947 $\pm$ 0.005} & \del{0.956 $\pm$ 0.015} \add{0.916 $\pm$ 0.015} \\
\hline
\multirow{2}{*}{9 features}
& ROC & \del{0.941 $\pm$ 0.015} \add{0.951 $\pm$ 0.003} & \del{0.964 $\pm$ 0.011} \add{0.973 $\pm$ 0.003} & \del{0.957 $\pm$ 0.011} \add{0.971 $\pm$ 0.003} & \del{0.956 $\pm$ 0.013} \add{0.961 $\pm$ 0.004} \\
& PR & \del{0.928 $\pm$ 0.024} \add{0.889 $\pm$ 0.009} & \del{0.950 $\pm$ 0.021} \add{0.923 $\pm$ 0.010} & \del{0.948 $\pm$ 0.021} \add{0.930 $\pm$ 0.009} & \del{0.937 $\pm$ 0.026} \add{0.888 $\pm$ 0.025} \\

\hline
\end{tabular}
\label{table_auc}
\end{table}

\begin{figure}[bt]
\centering
\includegraphics[width=0.97\linewidth]{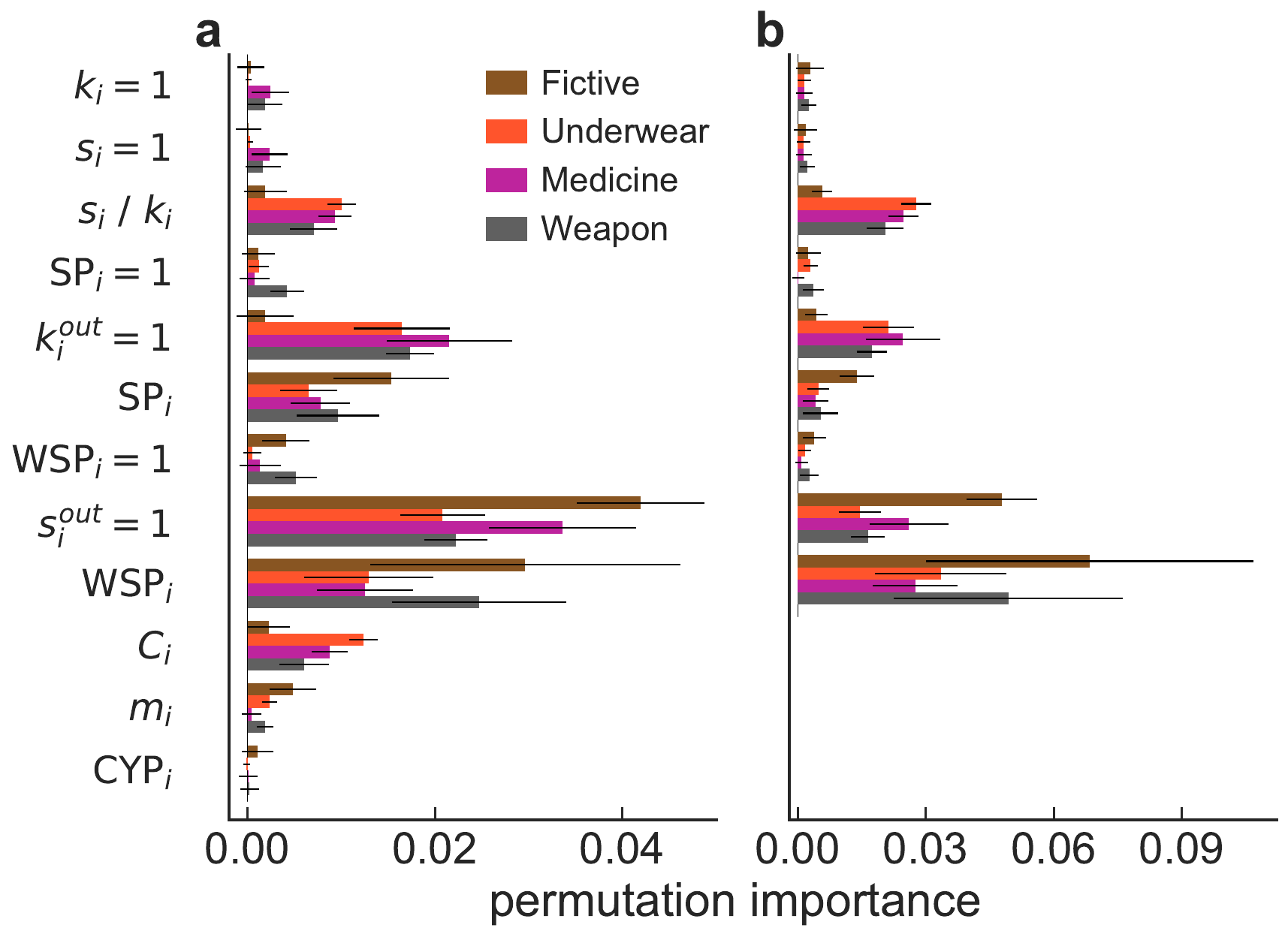}
\caption{{\bf Permutation importance of the features in the random forest 
classifier.}
(a) 12 features.
(b) 9 features.
The bars indicate the average over the 100 classifiers.
The error bars indicate standard deviation.}
\label{fig_feature}
\end{figure}

The importance of each feature in the classifier is shown in 
Fig.~\ref{fig_feature}a, separately for the different fraud types.
The importance of each feature is similar across the different types 
of fraud.
Figure~\ref{fig_feature}a indicates that 
the average number of transactions per neighbor (i.e., $s_i/k_i$), 
whether or not $k_i^\mathrm{out}=1$ (i.e., SP$_i=1/(k_i^\mathrm{in}+k_i^\mathrm{out})$),
whether or not $s_i^\mathrm{out}=1$ (i.e., WSP$_i=1/(s_i^\mathrm{in}+s_i^\mathrm{out})$),
and the weighted sell probability (i.e., WSP$_i$) are the four features of 
the highest importance.
Given the results of the descriptive statistics in the previous section, 
a small value of $s_i/k_i$, 
$k_i^\mathrm{out} \ne 1$, 
$s_i^\mathrm{out} \ne 1$, and
a moderate WSP$_i$ value
strongly suggest that the user may be fraudulent.


Figure~\ref{fig_feature}a also suggests that the features based on the 
triangles, i.e., $C_i$, $m_i$, and $\mathrm{CYP}_i$, are not strong 
contributors to the classifier's performance.
Because these features are the only ones that require the information about 
the connectivity between pairs of neighbors of the focal node, it is 
practically beneficial if one can realize a similar classification 
performance without using these features; 
then only the information on the connectivity of the focal users is required. 
To explore this possibility, we constructed the random forest classifier 
using the nine out of the twelve features that do not require the 
connectivity between neighbors of the focal node.
The mean AUC values for the ROC and PR curves are shown in 
Table~\ref{table_auc}.
We find that, despite some reduction in the performance scores relative 
to the case of the classifier using all the 12 features, the AUC values
with the nine features are still large, all exceeding \del{0.92} \add{0.88}.
The permutation importance of the nine features is shown in 
Fig.~\ref{fig_feature}b.
The results are similar to those when all the 12 features are used, 
although the importance of WSP$_i$ considerably increased in 
the case of the nine features (Fig.~\ref{fig_feature}a).

More than half of the normal users have $k_i=1$, and there are few 
fraudulent users with $k_i=1$ in each fraud category 
(Table~\ref{table_summary}).
The classification between the normal and fraudulent users may be an easy 
problem for this reason, leading to the large AUC values. 
To exclude this possibility, we carried out a classification test for 
the subdata in which the normal and fraudulent users with $k_i=1$ were
excluded, leaving 412 normal users and a similar number of fraudulent 
users in each category (Table~\ref{table_summary}).
We did not carry out subsampling because the number of the negative and 
positive samples were similar.
Instead, we generated 100 different sets of train and test samples and 
built a classifier based on each set of train and test samples.
The AUC values when either 10 or 7 features (i.e., the features excluding 
whether or not $k_i=1$ and whether or not $s_i=1$) are used are shown in 
Table~\ref{table_auc_kge2}.
The table indicates that the AUC values are still competitively large 
while they are smaller than those when whether or not $k_i=1$ and whether 
or not $s_i=1$ are used as features (Table~\ref{table_auc}).

\begin{table}[b]
\centering
\caption{{\bf AUC values for the random forest classifiers excluding 
users with $k_i=1$.}
The average and standard deviation were calculated based on the 100 
classifiers.}
\begin{tabular}{rrC{2.08cm}C{2.08cm}C{2.08cm}C{2.08cm}}
& & Fictive & Underwear & Medicine & Weapon \\
\hline
\multirow{2}{*}{10 features} & ROC & 0.925 $\pm$ 0.016 & 0.950 $\pm$ 0.013 & 0.954 $\pm$ 0.012 & 0.916 $\pm$ 0.019 \\
& PR & 0.923 $\pm$ 0.019 & 0.950 $\pm$ 0.018 & 0.954 $\pm$ 0.016 & 0.911 $\pm$ 0.023 \\
\hline
\multirow{2}{*}{7 features}& ROC & 0.886 $\pm$ 0.020 & 0.921 $\pm$ 0.015 & 0.933 $\pm$ 0.014 & 0.899 $\pm$ 0.020 \\
& PR & 0.874 $\pm$ 0.027 & 0.901 $\pm$ 0.021 & 0.928 $\pm$ 0.019 & 0.880 $\pm$ 0.028 \\
\hline
\end{tabular}
\label{table_auc_kge2}
\end{table}

\section*{Discussion}
We showed that a random forest classifier using network features of users 
distinguished different types of fraudulent users from normal users with 
approximately \del{0.95}\add{0.91}--0.98 in terms of the AUC. 
We only used the information about local transaction networks centered 
around focal users to synthesize their features.
We did so because it is better in practice not to demand the information 
about global transaction networks due to the large number of users.
It should be noted that AUC values of \del{$\approx 0.93$--0.96} \add{$\approx 0.88$--0.97} was also realized 
when we only used the information about the connectivity of the focal user, 
not the connectivity between the neighbors of the focal user.
This result has a practical advantage when the present fraud-detection 
method is implemented online because it allows one to classify users with 
a smaller amount of data per user.

The random forest classifier is an arbitrary choice. 
One can alternatively use a different linear or nonlinear classifier to 
pursue a higher classification performance.
This is left as future work.
Other future tasks include the generalizability of the present results to 
different types of fraudulent transactions, such as resale tickets, 
pornography, and stolen items, and to different platforms.
In particular, if a classifier trained with test samples from fraudulent 
users of a particular type and normal users is effective at detecting 
different types of fraud, the classifier will also be potentially 
useful for detecting unknown types of fraudulent transactions.
It is also a potentially relevant question to assess the classification 
performance when one pools different types of fraud as a single 
positive category to train a classifier.

Prior network-based fraud detection has employed either global or local 
network properties to characterize nodes.
Global network properties refer to those that require the structure of the 
entire network for calculating a quantity for individual nodes, such as 
the connected component
\cite{subelj2011, savage2017, wang2018}, 
betweenness centrality\cite{subelj2011, drezewski2015, colladon2017}, 
user's suspiciousness determined by belief propagation
\cite{chau2006, pandit2007, akoglu2013, bangcharoensap2015, 
vlasselaer2015, vlasselaer2016, hu2017, li2017}, 
dense subgraphs including the case of communities
\cite{subelj2011, bhat2013, ferrara2014, jiang2014, hooi2016, liu2016, shchur2018,},
and $k$-core\cite{wang2008, rasheed2018}.
Although many of these methods have accrued a high classification 
performance, they require the information about the entire network.
Obtaining such data may be difficult when the network is large or rapidly
evolving over time, thus potentially compromising the computation 
speed, memory requirement, and the accuracy of the information on the 
nodes and edges.
Alternatively, other methods employed local network properties such as 
the degree including the case of directed and/or weighted networks
\cite{chau2006, akoglu2010, subelj2011, yanchun2011, bhat2013, 
bangcharoensap2015, drezewski2015, monamo2016, vlasselaer2016, colladon2017} 
and the abundance of triangles and quadrangles
\cite{monamo2016, vlasselaer2016}.
The use of local network properties may be advantageous in industrial
contexts, particularly to test sampled users, because 
local quantities can be rapidly calculated given a seed node.
Another reason for which we focused on local properties was that we could 
not obtain the global network structure for computational reasons.
It should be noted that, while the use of global network properties in 
addition to local ones may improve the classification accuracy
\cite{bhat2013}, the present local method attained a similar 
classification performance to those based on global network properties,
i.e., 0.880--0.986 in terms of the ROC AUC
\cite{subelj2011, vlasselaer2015, vlasselaer2016, hu2017, li2017, savage2017}.

A prior study using data from the same marketplace, Mercari, aimed to 
distinguish between desirable non-professional frequent sellers and 
undesirable professional sellers\cite{yamamoto2019}.
The authors used information about user 
profiles, item descriptions, and other behavioral data such as the number 
of purchases per day. 
In contrast, we focused on local network features of the users (while a 
quantity similar to WSP$_i$ was used as a feature in 
Ref.\cite{yamamoto2019}).
In addition, we used specific types of fraudulent transactions, whereas
the authors of Ref.\cite{yamamoto2019} focused on  problematic 
transactions as a single broad category.
\add{How the present results generalize to different categorizations of 
fraudulent transactions, the platform's different data such as their US 
market data, and similar data obtained from other online marketplaces 
is unknown.}
Combining network and non-network features may realize a better 
classification performance
\del{, which also warrants future work}.
\add{Furthermore, using the information about the time of the transactions 
may also yield better classification.
Using the time information allows us to ask new questions such as 
prediction of users' behavior.
These topics warrant future work.}

\section*{Declarations}
\subsection*{Abbreviations}
C2C: consumer-to-consumer;
ROC: receiver operating characteristic;
PR: precision-recall;
AUC: area under the curve

\subsection*{Ethics}
Mercari, Inc. approved the use of the data for the present study 
under the condition that the data were hashed and only released to 
the collaborators of the project (i.e., the first and last authors, 
because the second and third authors are employees of the company).

\subsection*{Availability of data and materials}
Mercari, Inc. approved the use of the data for the present study 
under the condition that the data were hashed and only released to 
the collaborators of the project (i.e., the first and last authors, 
because the second and third authors are employees of the company). 
The figures and tables of the present paper are summary statistics of 
the data and not sufficient on their own for others to replicate the 
results of the present study. 
Although the data have been hashed, the company cannot share the data 
with the public. 
This is because, if anybody traces the transaction data on the 
Mercari's web platform and checks them against the hashed data, 
that person would be able to identify individual users including their 
private information. 
Therefore, hashing/anonymizing does not help to guarantee the users' 
privacy. 
Any bona fide researcher could approach the company (Shunya Kimura: 
kimuras@mercari.com and Ryusuke Chiba: metalunk@mercari.com) to seek 
access to the complete dataset. 
However, for the aforementioned reasons, such an attempt is unlikely 
to be successful. 

The users were made aware that their data may be used for the present 
research because the Mercari's terms of use (in Japanese only: 
https://www.mercari.com/jp/tos/), Article 20, Term 2, states that 
their data can be used for research by the company and by those who 
the company permits.

\subsection*{Competing interests}
The second and third authors are employees of the company that 
provided the data analysed in the present manuscript. 
However, this fact does not cause any conflict of interest because 
the analyses, results and their interpretation are free of any bias 
towards the merit of the company.

\subsection*{Funding}
The authors acknowledge financial support by Mercari, Inc.
S. Kodate was supported in part by the Top Global University Project 
from the Ministry of Education, Culture, Sports, Science and 
Technology (MEXT) of Japan. 

\subsection*{Authors' contributions}
S. Kodate analyzed data, developed methodology, visualized the 
results, and drafted the manuscript; 
RC curated data and critically revised the manuscript; 
S. Kimura coordinated the study, acquired funding, and critically 
revised the manuscript; 
NM coordinated the study, acquired funding, developed methodology, 
drafted the manuscript. 
All authors gave final approval for publication and agreed to be held 
accountable for the work performed therein.

\subsection*{Acknowledgments}
This work was carried out using the computational facilities of the 
Advanced Computing Research Centre, University of Bristol.

\bibliography{mercari_paper_ans}
\newpage

\end{document}